\documentclass[12pt]{iopart}

\usepackage{graphicx}
\usepackage{cite}
\usepackage{amssymb}

\begin{document}

\title[]{Extending the Near-infrared Band-edge Absorption Spectrum of Silicon by Proximity to a 2D Semiconductor}

\author{Valerio~Apicella$^{1,2,\dagger}$, Teslim~Ayinde~Fasasi$^{1,\dagger}$, Antonio Ruotolo$^{1,*}$}
\address{$^1$Department of Materials Science and Engineering, City University of Hong Kong, Kowloon, Hong Kong SAR, China}
\address{$^2$Department of Engineering, University of Sannio, Benevento 82100, Italy}
\address{$^\dagger$These authors contributed equally }
\address{$^*$ Corresponding Author}
\ead{aruotolo@cityu.edu.hk}
\vspace{10pt}
\begin{indented}
	\item[]\today
\end{indented}

\begin{abstract}
 Because of its low-cost, silicon is the standard material for photovoltaic conversion. Yet, its band-edge absorption spectrum is narrower than the spectrum of the solar radiation, which reduces its conversion efficiency. In this paper, it is shown that the spectrum of absorbance of silicon can be extended to longer wavelengths by proximity to a two-dimensional (2D) semiconductor. Photo-induced Hall effect, together with standard absorption spectroscopy, was employed to estimate the increase of efficiency of absorbance of a 2D-platinum-diselenide/intrinsic-silicon bilayer. The bilayer shows a significantly higher absorption in the infrared as compared to the single films. Moreover, an overall increase of absorption efficiency by a factor twenty was measured in the entire spectrum of light of a halogen lamp. X-ray Photoelectron Spectroscopy (XPS) confirms that a reduction of the band-gap occurs in the silicon substrate at the interface between the two semiconductors. The results are interpreted in the framework of band-gap narrowing due to hole-plasma confinement in the Si, induced by electron-confinement in the 2D film. Possible application of the effect in photo-voltaic cells is discussed.
\end{abstract}

\noindent{\it Keywords}: 2D semiconductors, photo-conversion, band-gap engineering

\section{Introduction}
The need for inexpensive renewable energy sources and the potentialities of harvesting the huge energy amount coming from the sun in a more efficient way, has been pushing both industry and research of last decades towards finding new solutions to increase the solar cells efficiency. 
Among the large variety of materials exploited in photovoltaics, silicon remains the most used one because of its abundance and its consequently very low-cost. Nevertheless, its band-gap of 1.12~eV represents one of the major issues limiting its performances in photovoltaic conversion \cite{tiedje1984limiting,richter2013reassessment}. It sets the edge of the absorption spectrum at about 1107~nm \cite{shalav2005application}. As a consequence, the largest amount of the near-infrared solar spectrum content is lost and does not take part to the energy conversion process. Different solutions have been proposed in literature to enhance the absorption of the infrared light. The possibility of coupling more semiconductors in multi-gap \cite{marti1996limiting}, or multi-junction \cite{tobias2002ideal}, tandem solar cells is an example. In particular, two or more solar cells with different absorption characteristics are series-connected together thorough tunnel junctions with the aim of covering a wider part of the solar spectrum. Recent examples of tandem solar cells make use of polymers \cite{you2013polymer} and organic cells \cite{meng2018organic}.
\\In the same direction goes the exploitation of thin films to improve the infrared absorption \cite{lee2017review}. First examples are given in refs.~\cite{devaney1990structure,brabec2002low} where a CuInGaSe$_2$ and a low band-gap polymer were proposed, respectively. More recently, the usage of quantum dots \cite{yan2010large,tang2011infrared} and nanowires \cite{lin2009optical,kelzenberg2010enhanced} was proposed for infrared photovoltaics.
\\After graphene was discovered in 2004 \cite{novoselov2004electric}, new possibilities arose in developing photovoltaic devices \cite{liu2008organic,chien2015graphene}, leading also to the usage of other two-dimensional (2D) materials \cite{liu2015functionalized}. Among the latter class of semiconductors, most promising for infrared absorption and detection seem to be the transition metal dichalcogenides (TMDs) \cite{wang2012electronics,bernardi2013extraordinary,jariwala2014emerging,mak2016photonics,tan2018emerging}.
\\In this paper, a simple method to extend the near-infrared band-edge absorption spectrum of silicon (Si) is presented. In particular, 2D platinim diselenide (PtSe$_2$), a narrow band-gap TMD semiconductor, is exploited. As explained in the following, the large difference in band-gap between the two semiconductors results in an electron confinement in the 2D layer, which induces a hole-plasma in the Si. The concentration of holes decays as the inverse of the distance from the interface, resulting in a gradual narrowing of the band-gap and, therefore an increase in both the absorbance efficiency and absorbance spectrum. The high resistivity of the intrinsic silicon favours the confinement of the hole-plasma near the interface. Photo-induced Hall effect \cite{li2018photo}, together with standard spectroscopy, allowed us to estimate the increase of photo-conversion efficiency without having to resort to electrical measurements, which would require the use of doped semiconductors. The bilayer behaves as a single, engineered semiconductor with an overall increase of 20 times in absorbance, partially due to an increase of absorbance in the infrared range, which takes place near the interface in the Si, not in the 2D semiconductor. Our interpretation is supported by angle-resolved X-ray Photoelectron Spectroscopy (XPS) measurements, which confirm that a reduction of the band-gap exists in the Si near the interface.

\section{Results}
The two-dimensional semiconductor used in the present work is a two/three atomic layer platinum diselenide (PtSe$_2$) with 1.8~nm thickness. It is a transition metal dichalcogenide which has been demonstrated to undergo a transition from metal to semiconductor by reducing its thickness from bulk to two-dimensional layers \cite{ciarrocchi2018thickness}. The system of interest here is an $1 \times 1$~cm$^2$ PtSe$_2$/Si bilayer, where the Si substrate was chosen to be intrinsic Si (i-Si) for reasons that will be explained in the following. An $1 \times 1$~cm$^2$ PtSe$_2/\alpha$-Al$_2$O$_3$ bilayer was used as reference in optical measurements, while two more reference samples, namely Pt/PtSe$_2$/$\alpha$-Al$_2$O$_3$ and Pt/Si, were used as reference for photo-induced Hall measurements. 

\subsection{Raman Spectroscopy}
\begin{figure}[!ht]
	\centering
	\includegraphics[width=0.5\columnwidth]{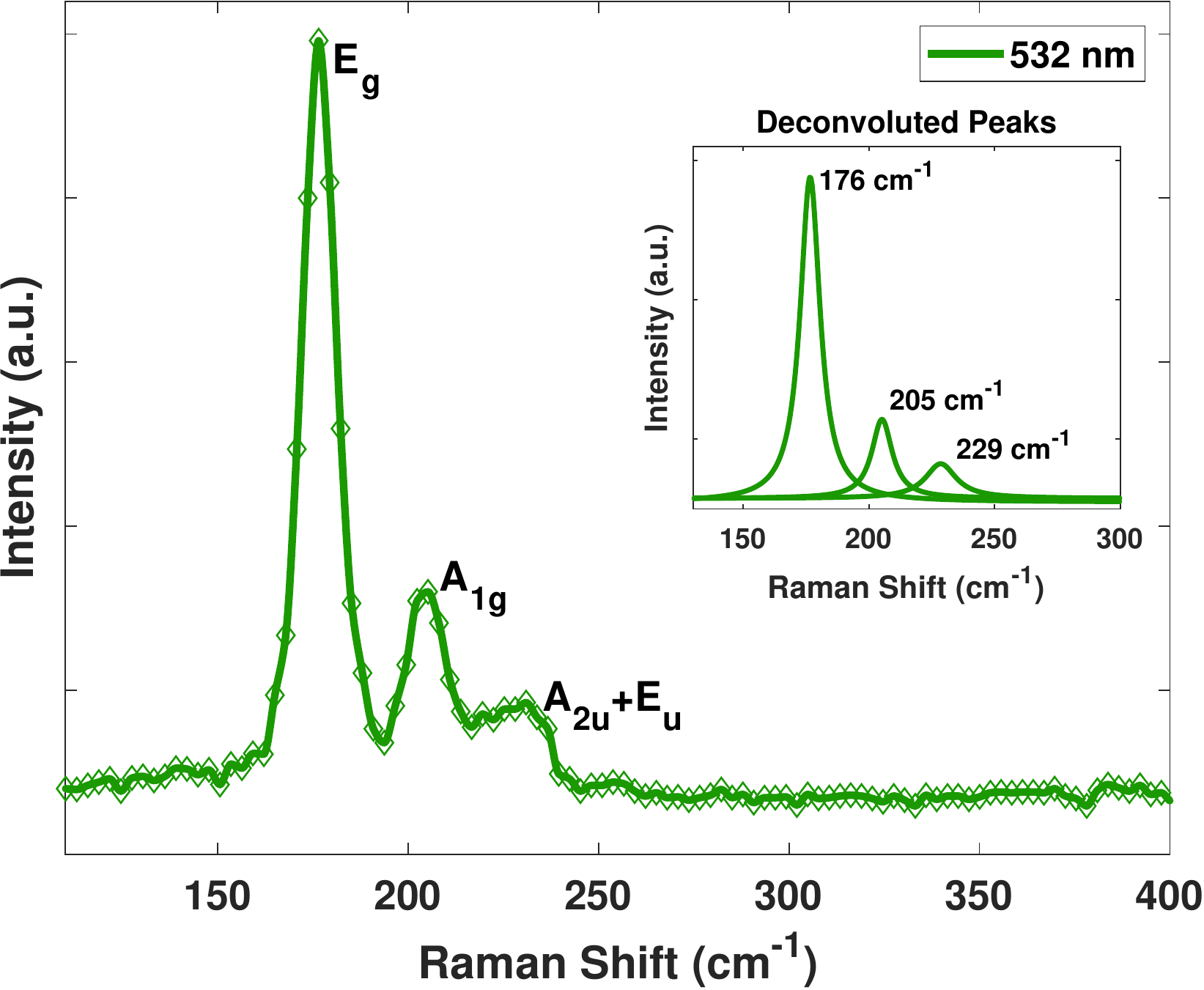}
	\caption{Raman spectrum of the 2D PtSe$_2$ sample on the sapphire substrate at 532~nm laser. The deconvoluted peaks are shown in the inset.}
	\label{fig: Raman_2D_sapphire}
\end{figure}

The measured Raman spectrum of the 2D film is shown in \figurename{~\ref{fig: Raman_2D_sapphire}}.  The measurement was carried out on the PtSe$_2/\alpha$-Al$_2$O$_3$ sample, in order to avoid the high intensity signal from the Si. The spectrum is consistent with the one expected for the 2D semiconductor \cite{o2016raman}.
In particular, two main peaks at 176~cm$^{-1}$ and 205~cm$^{-1}$ and one with lower intensity at 229~cm$^{-1}$ can be observed. The first two are assigned to E\textsubscript{g} and A\textsubscript{1g} modes, respectively. The third one corresponds to the overlapping of A\textsubscript{2u} and E\textsubscript{u} modes. As discussed in ref.~\cite{o2016raman} the latter peak disappears for thickness of about 5 nm. Furthermore, the higher its relative intensity, with respect to the more prominent ones, the thinner the PtSe$_2$ layer. As a consequence, the Raman spectrum in \figurename{~\ref{fig: Raman_2D_sapphire}} confirms the two-dimensional nature of the PtSe$_2$ layer under test.

\subsection{Fourier Transform Infrared Spectroscopy}
\begin{figure*}[!ht]
	\centering
	\includegraphics[width=0.49\columnwidth]{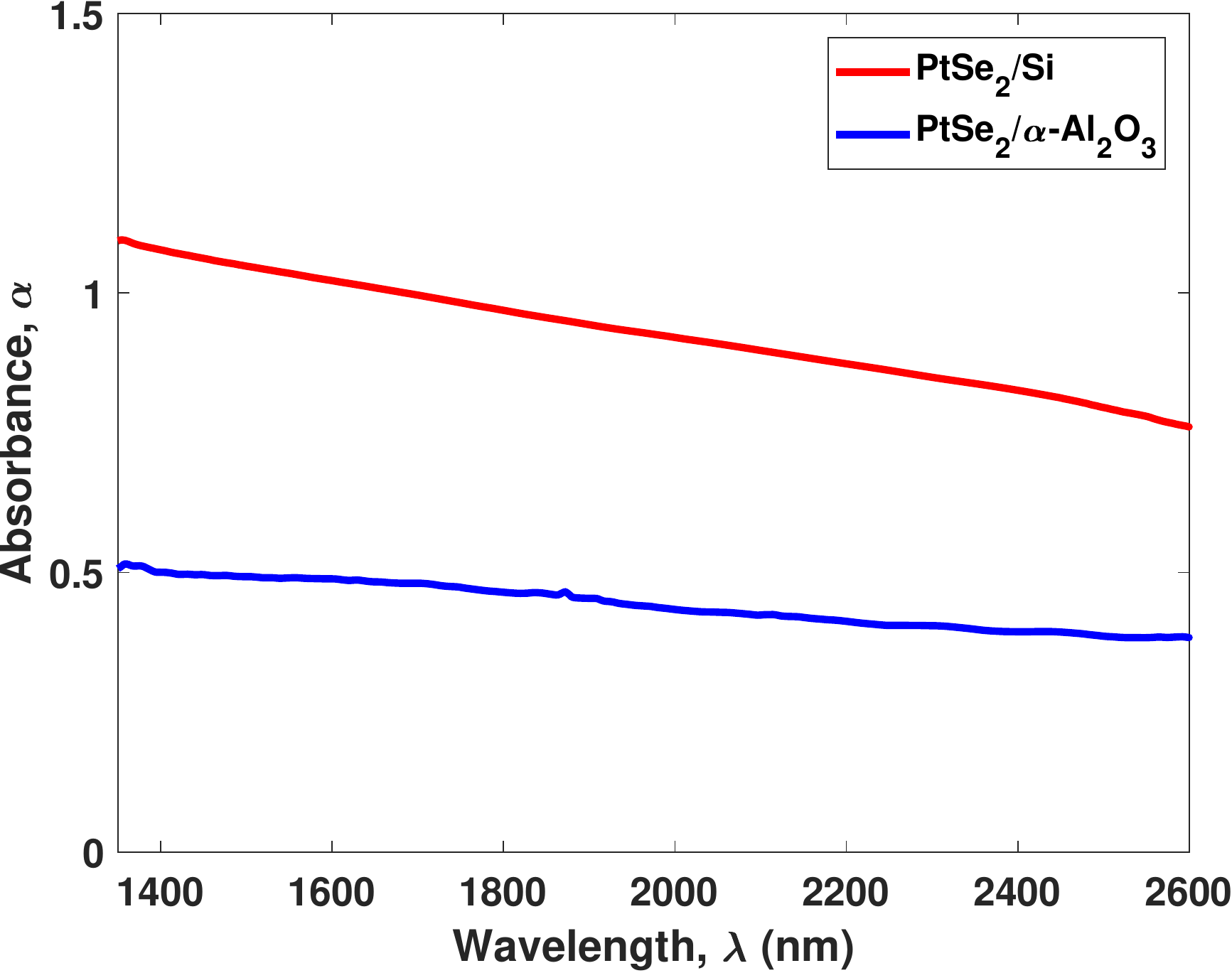}
	\caption{FTIR absorbance spectrum of PtSe$_2/\alpha$-Al$_2$O$_3$ and PtSe$_2$/Si.}
	\label{fig: FTIR}
\end{figure*}

\begin{figure}[!ht]
	\centering
	\includegraphics[width=0.49\columnwidth]{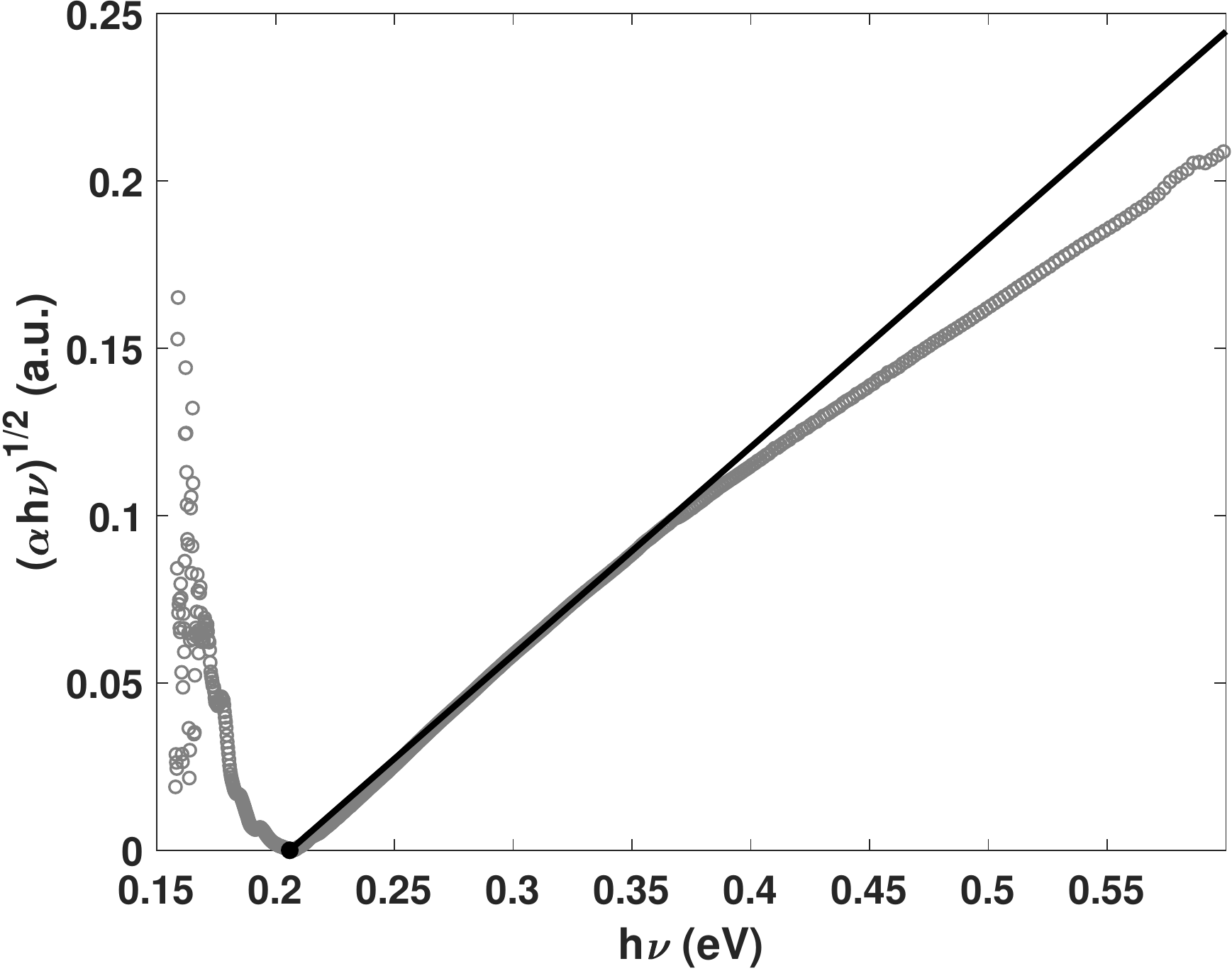}
	\caption{Tauc plot calculated from the FTIR absorbance spectrum of the 2D PtSe$_2$ sample on the sapphire substrate. The band-gap of the 2D layer is estimated to be approximately equal to 0.21~eV.}
	\label{fig: tauc_plots}
\end{figure}

In order to investigate how the proximity to the 2D semiconductor affects the optical behavior of the silicon substrate, \emph{Fourier Transform Infrared} (FTIR) spectroscopy measurements were performed on the two samples. Since the bandgap of Si is $E_{g,Si} = 1.12$~eV, corresponding to a cut-off absorption wavelength of $\lambda = 1107$~nm, Si is expected to be completely transparent in the infrared. On the contrary, in \figurename{~\ref{fig: FTIR}} one can see that a significant absorbance exists in the near-infrared, beyond 1107~nm in the PtSe$_2$/Si sample. The absorbance is not negligible up to 6000~nm, as shown in the full absorbance spectrum in Figure S1 (see Supporting Information). The same measurement carried out on the PtSe$_2/\alpha$-Al$_2$O$_3$ reference sample shows that the absorbance in the 2D film in the near-infrared is lower with respect to the first sample. The increase of absorbance in the reference sample for wavelengths longer than 5000~nm (see Figure S1, Supporting Information) can be ascribed to the absorbance of the sapphire substrate \cite{thomas1988infrared,druy1990fourier}, and is of no interest in the present framework.
The band-gap of the 2D semiconductor was estimated through the \emph{Tauc} analysis \cite{tauc1966optical} of the absorbance spectrum of the PtSe$_2/\alpha$-Al$_2$O$_3$ reference sample.  
The Tauc plot of the absorbance spectrum is shown in \figurename{~\ref{fig: tauc_plots}}. The noise at low photon energies is due to the aforementioned absorbance of the sapphire for wavelengths longer than 5000~nm. Then, a clear change of slope is observed for photon energies near the expected band-gap value. The absorbance depends on the difference between photon energy and band-gap, as $(\alpha h\nu)^{1/n} \propto (h\nu-E_g)$, where $n=2$ for indirect band-gap semiconductors \cite{viezbicke2015evaluation}, such as in the case of 2D PtSe$_2$ \cite{kandemir2018structural}. As a consequence, a bandgap $E_{g,2D} = 0.21$~eV was estimated by using the intercept to the $x-$axis (see \figurename{~\ref{fig: tauc_plots}}). This value is in agreement with ref.~\cite{wang2015monolayer}, in which a band-gap of 0.21~eV was reported for monolayer PtSe$_2$.

\subsection{Photo-induced Hall effect measurements}
\begin{figure}[!ht]
	\centering
	\includegraphics[width=1\columnwidth]{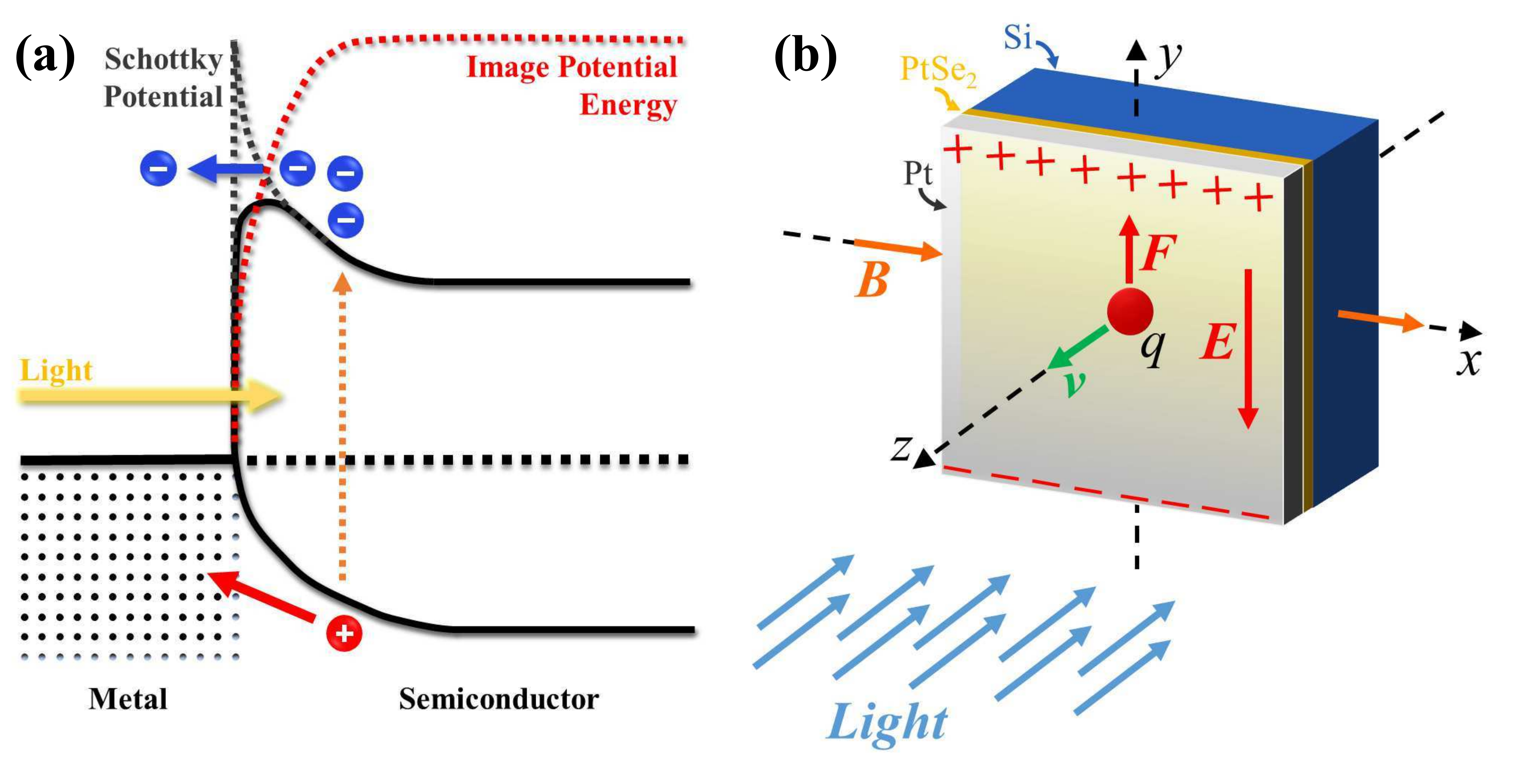}
	\caption{Photo-induced Hall effect working principle: (a) barrier potential rounding-off due to image force effect; (b) photo-induced Hall voltage generation in the Pt/PtSe$_2$/Si sample.}
	\label{fig: photo_hall_principle}
\end{figure}

\begin{figure*}[!ht]
	\centering
	\includegraphics[width=1\columnwidth]{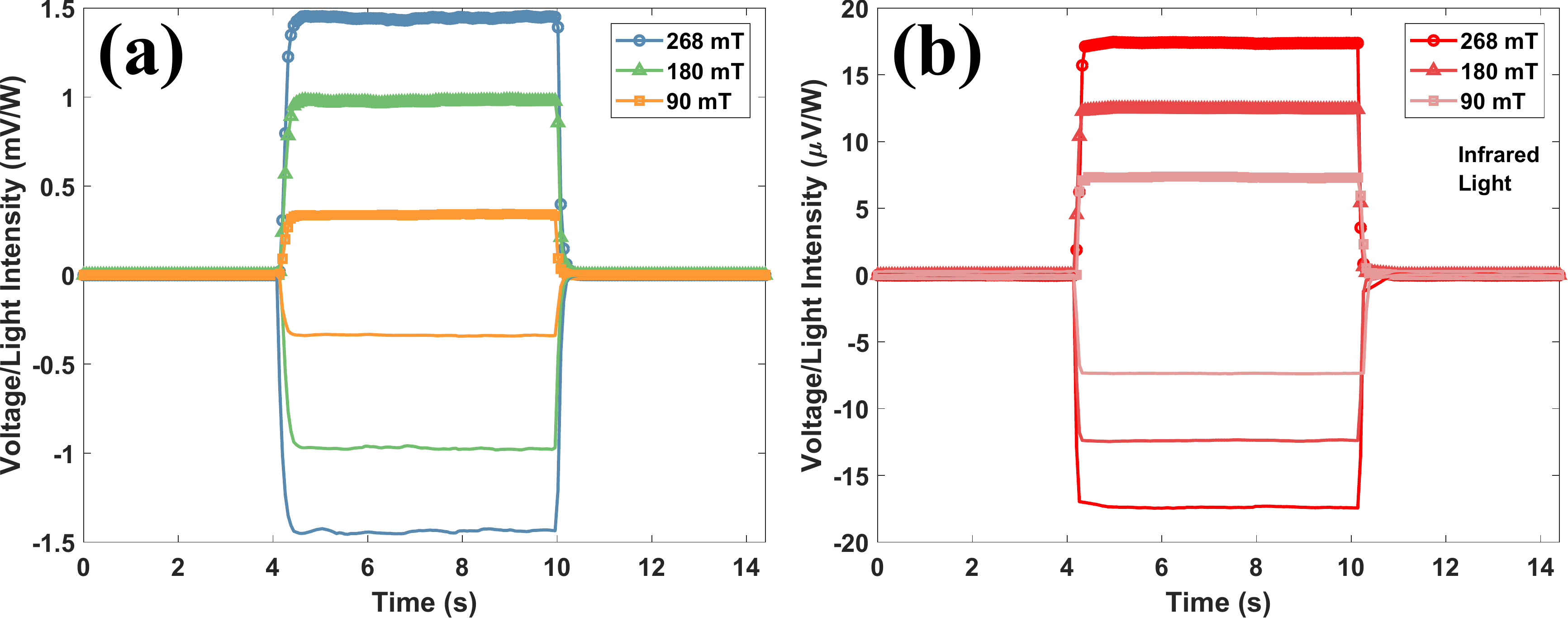}
	\caption{Photo-induced Hall effect measurements on the Pt/PtSe$_2$/Si sample in the full halogen lamp spectrum (a) and infrared light (b). Both positive and negative magnetic fields were applied and represented with marked and unmarked lines, respectively.}
	\label{fig: photo_hall_2D}
\end{figure*}

\begin{figure*}[!ht]
	\centering
	\includegraphics[width=1\columnwidth]{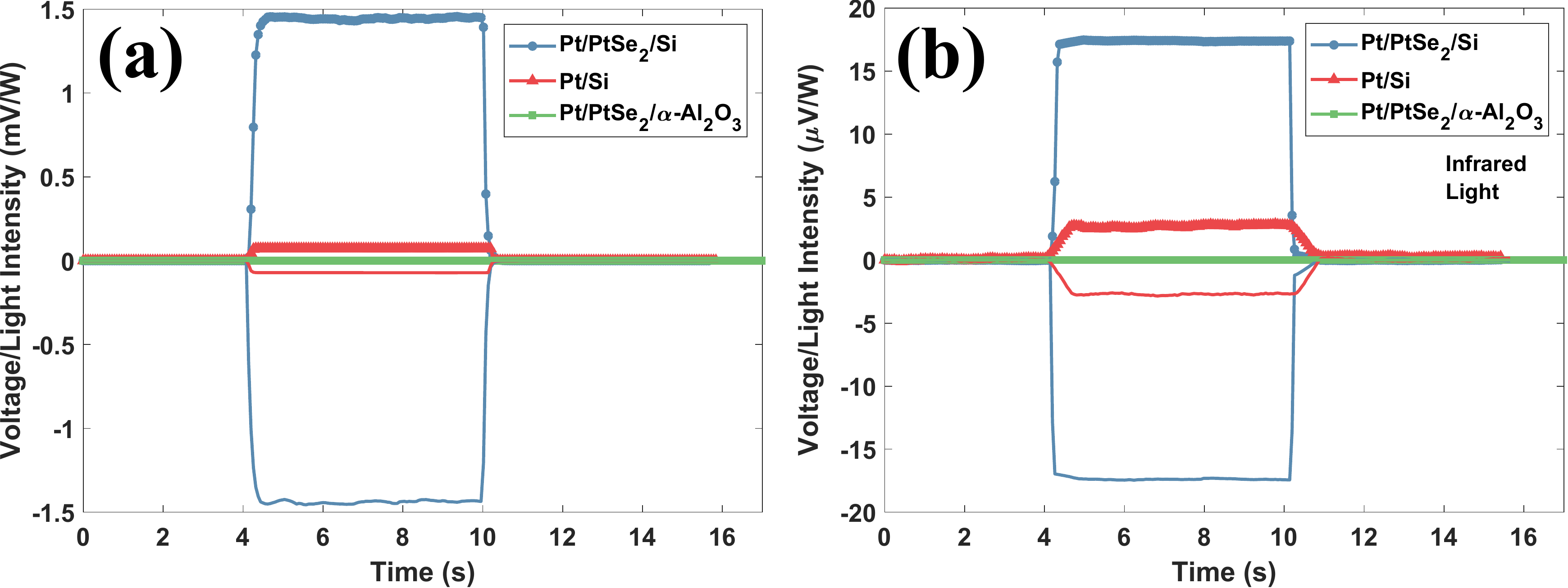}
	\caption{Comparison of photo-Hall effect for Pt/PtSe$_2$/Si, Pt/Si and Pt/PtSe$_2$/$\alpha$-Al$_2$O$_3$ in full halogen spectrum (a) and infrared light (b). Both positive and negative magnetic fields with 268~mT amplitude were applied and represented with marked and unmarked lines, respectively.}
	\label{fig: photo_hall_comparison}
\end{figure*}

Photo-induced Hall effect \cite{li2018photo} offers the possibility to characterize the optical behavior of a semiconductor without flowing of a net-charge current. As will be discussed in the next section, this is important to exclude band-gap narrowing due to excitation and injection of high-density carriers through the 2D/Si interface \cite{del1987modelling}.
The working principle is briefly explained in the following with the help of \figurename{~\ref{fig: photo_hall_principle}}. A metal with high work-function, such as platinum (Pt) is deposited on the semiconductor or, in this case, on the semiconductor hetero-structure. The thickness of the metal is chosen in such a way that light can penetrate without significant attenuation for wavelengths as long as infrared radiations. In this experiment the thickness of the Pt was 3~nm. The metal forms a Schottky barrier to the semiconductor. As photons are converted into electron-hole pairs, while holes can easily be neutralized by electrons in the metal, electrons are confined near the interface, if the semiconductor is highly resistive. This results in a rounding-off of the barrier due to image force effect \cite{chang1970carrier}. The reduction of the barrier potential allows electron to be injected into the metal, as schematically shown in \figurename{~\ref{fig: photo_hall_principle}a}. If a magnetic field is applied in the sample plane, an open-circuit voltage appears that is transverse to the metal (\figurename{~\ref{fig: photo_hall_principle}b}). At equilibrium, the current of the holes and the current of the electrons are equal and no net-current flows, while the current densities are not uniform because of the Lorentz's forces exerted on the carriers. 
\figurename{~\ref{fig: photo_hall_2D}a} shows the open-circuit voltage detected on the Pt/PtSe$_2$/Si when illuminated by a halogen lamp and for different values of the applied magnetic field. \figurename{~\ref{fig: photo_hall_2D}b} shows the same measurement when a long-pass, infrared filter with cut-off wavelength of 1250~nm (which is safely longer than 1107, $\it{i.e.}$ the absorption spectrum of silicon), is placed in front of the sample. In both cases, the Hall voltage was normalized to the measured light intensity. One can clearly see that photo-conversion occurs in the near-infrared, in agreement with the FTIR measurements.
In order to understand where absorption of infrared occurs we prepared two additional reference samples, Pt/PtSe$_2$/$\alpha$-Al$_2$O$_3$ and Pt/Si for photo-induced Hall effect. In \figurename{~\ref{fig: photo_hall_comparison}a} and \figurename{~\ref{fig: photo_hall_comparison}b} we compare the photo-induced Hall effect measured on the three samples at  a fixed value of the magnetic field. 
In agreement with the FTIR measurements, absorption of infrared light is highly enhanced in the Pt/PtSe$_2$/Si bilayer as compared to the case of single semiconductors.
\\Finally, photo-induced Hall measurements with daylight blue and red filters were carried out and shown in Figure S2 (see Supporting Information). By comparing the relative intensities of the signals obtained over the two samples, it is straightforward to note that a shift in the absorbance peak to longer wavelengths is obtained in the Pt/PtSe$_2$/Si sample with respect to the Pt/Si one. This demonstrates that the 20 times increase of the overall absorbance mainly occurs in the red and infrared wavelength range.

\subsection{X-ray Photoelectron Spectroscopy}
\begin{figure}[!ht]
	\centering
	\includegraphics[width=0.49\columnwidth]{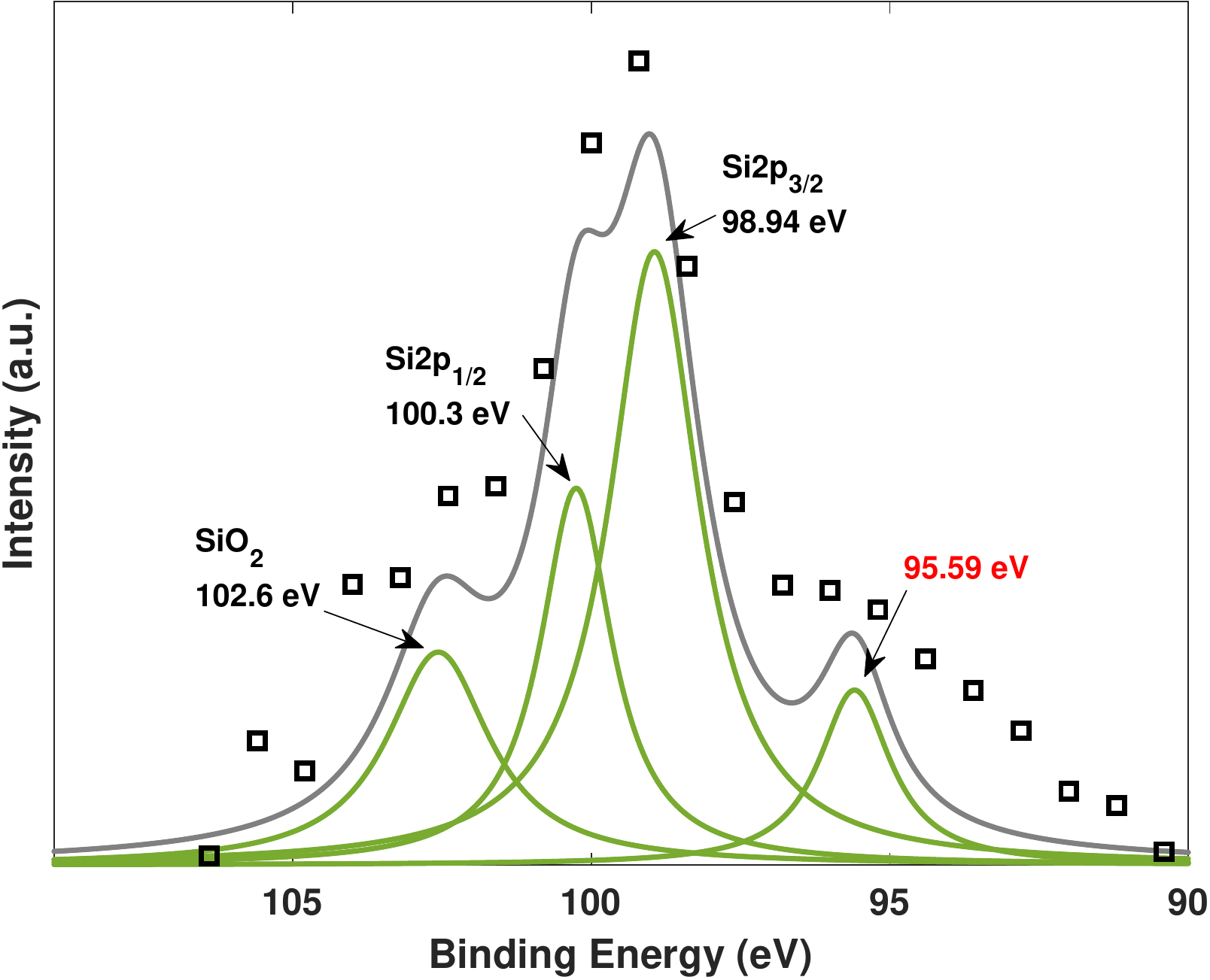}
	\caption{Deconvolution of the Si XPS peaks. The two peaks at 98.94~eV and 100.3~eV correspond to Si2p$_{3/2}$ and Si2p$_{1/2}$, respectively. The peak at 102.6~eV corresponds to SiO$_2$. The further peak at lower energy (95.59~eV) shows that a Si band-gap narrowing occurs near the interface.}
	\label{fig: XPS-Si}
\end{figure}

X-ray Photoelectron Spectroscopy (XPS) measurements were carried out on the PtSe$_2$/Si sample. The angle of the beam with respect to the sample was changed until the peaks of the Si under the 2D film could be resolved. The full spectrum is shown in Figure S3 (see Supporting Information). The Si peaks are shown in \figurename{~\ref{fig: XPS-Si}}. One can clearly observe the presence of four peaks that are better resolved by Lorentzian deconvolution. The two peaks at 98.94~eV and 100.3~eV are the expected 2p$_{3/2}$ and 2p$_{1/2}$ peaks of Si \cite{cerofolini2003si}. The peak at higher binding energy 102.6 eV can be attributed to residues of SiO$_2$ \cite{xpsHB} due to the transfer process of the 2D material on the silicon substrate (see Materials and Methods). Surprisingly, a forth peak can be resolved at lower binding energies, at around 95.59~eV. As discussed in the following section, this peak suggests a narrowing of the Si band-gap near the interface.

\section{Discussion}
The results presented in the previous section suggest a profound change of the band-structure of the Si near the interface on a mesoscopic scale. In particular, they suggest a band-gap narrowing (BGN), which results in both an increase of the absorption efficiency in the visible range and an extension of the absorption spectrum into the near infrared. A BGN of the order of hundreds of mV in Si can only be ascribed to the presence of an electron-hole plasma \cite{persson2000plasma}. In brief, the band-gap is by definition the minimum energy required to generate an unbound electron-hole pair. The distance between the electron and the hole in the crystal is such that electrostatic attraction between the two particles can be neglected. On the contrary, excitons in a crystal can be excited by providing smaller energy. An exciton is a quasiparticle in which an electron and a hole are attracted to each other by the electrostatic Coulomb force. The existence of a plasma favours conversion of photons into excitons.  This requires less energy than that required to generate an unbound electron-hole pair. A reduction of the energy that a photon must possess to excite an electron-hole pair is equivalent to a reduction of the band-gap. 
\\The plasma can have three possible origins: i) heavy doping; ii) optical excitation of a large concentration of electron-hole pairs, usually through laser sources and iii) injection of high-density carriers through the interface of a biased junction, in which case the plasma, and therefore the BGN, is confined near the interface. Heavy doping in silicon has indeed been proven effective to extend its spectrum of absorption to the infrared \cite{balkanski1969infrared,schmid1981optical}. Of course, this possibility can here be excluded because the silicon used in this work is intrinsic. Excitation of a large concentration of electron-hole pairs can also be excluded, because in our experiment, unlike the Pt/PtSe$_2$/Si sample, the Pt/Si sample does not absorb infrared light, under the same experimental conditions. 
Last, injection of high-density carriers through the interface can be ruled out because the FTIR in \figurename{~\ref{fig: FTIR}} shows that a BGN of the bilayer exists under no electrical bias. Besides, in photo-induced Hall effect, no net-current flows through the hetero-junction. One must conclude that in our system the existence of a plasma confined near the interface in the Si must be due to the proximity to the 2D film.% 
\begin{figure*}[!ht]
	\centering
	\includegraphics[width=1\textwidth]{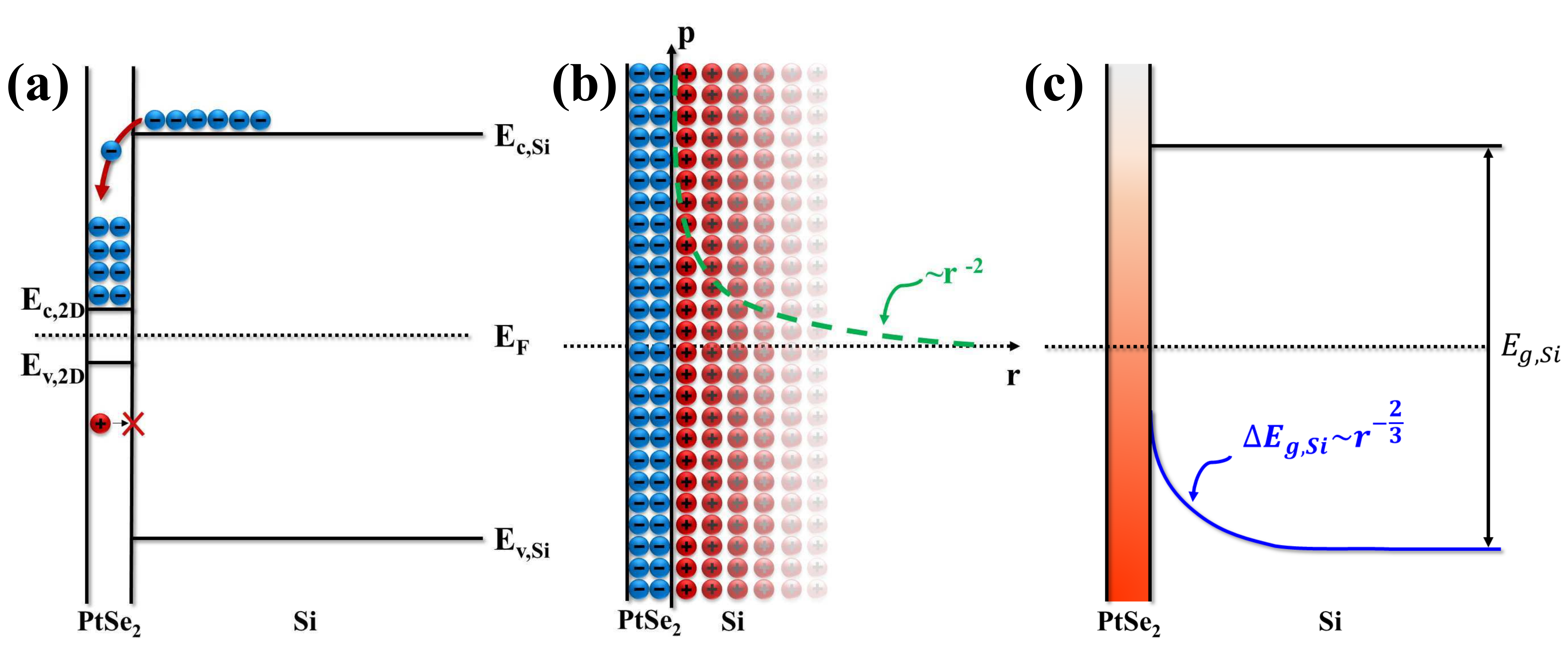}
	\caption{Schematic explanation of the BGN near the interface: (a) electron in the Si conduction band migrate to the 2D semiconductor; (b) an electron-hole plasma appears near the interface due to the electron confinement in the 2D layer; (c) the upward bending of the silicon valence band induces the band-gap narrowing near the interface.}
	\label{fig: sketch_disc}
\end{figure*}
As the Si substrate is placed in intimate contact with the 2D semiconductor, electrons in the conduction band (CB) of the Si see a large density of energy states available across the interface and migrate (see \figurename{~\ref{fig: sketch_disc}a}). On the contrary, holes from the 2D film see a large potential barrier and cannot migrate into the Si. The 2D behaves as a potential well for the electrons. As electrons are confined in a two/three atomic layers thick film, Coulomb interaction is not negligible. On one side, confinement of electrons in a 2D semiconductor is well known to induce a change in the band-structure of the 2D \cite{alidoust2014observation}, which we shall not discuss further here because the effect we observe is in the bulk semiconductor. From simple electrostatic considerations, an electron plasma confined in the 2D must induce a hole plasma in the Si, in which the charge density $p$ decays as $p \sim r^{-2}$, where $r$ is the distance from the interface (see \figurename{~\ref{fig: sketch_disc}b}). The BGN is known to increases as $\sim p^{1/3}$ \cite{jain1991simple}. Therefore, $\Delta E_{g,Si} \sim r^{-2/3}$  (see \figurename{~\ref{fig: sketch_disc}c}), where the singularity in $r = 0$ is due to the fact that the discrete nature of matter has not been taken into account. As a hole-plasma exists in the Si, near the interface, an upward bending of the valence band (VB) is at the origin of the BGN. This scenario is supported by the XPS spectrum shown in \figurename{~\ref{fig: XPS-Si}}. In this measurement, the low energy peak at 95.59 eV can not be assigned to SiO$_2$ nor to Pt and Se in the 2D semiconductor. Peaks related to the latter two elements are located at much lower binding energies and are shown in Figure S4 and Figure S5 (see Supporting Information), respectively. As a consequence, the peak at 95.59 eV indicates that a reduction of the band-gap occurs in the Si substrate near the interface with the 2D layer. This BGN can be regarded as a 2D/semiconductor version of the rounding-off of the CB edge in metal/semiconductor Schottky junctions due to confinement of electrons in the semiconductor near the interface (see \figurename{~\ref{fig: photo_hall_principle}a} and \figurename{~\ref{fig: sketch_disc}c}). Except, in 2D/semiconductor bilayers, electrons are confined in the 2D, which results into a bending of the VB, rather than the CB. 
It is interesting to notice that a similar bending of the valence band of a bulk semiconductor in proximity to a 2D layer was recently reported in ref.~\cite{molle2016electron} and ascribed to electron confinement, although in the cited work only an electrical characterization of the bilayer was carried out and therefore no evidence of BGN was reported.  
Let us also point out that the effect we here observe should not be confused with the proximity effect due the penetration of
electronic wavefunctions of one semiconductor into the other, which is purely quantum-mechanical and completely negligible in semiconductors.
\\Finally, let us speculate on how this effect could be employed in a photo-voltaic cell. If the intrinsic silicon substrate were to be replaced by doped Si to work as the active layer of a photo-voltaic cell, we predict the BGN to decrease as the doping level increases. This is because the hole-plasma would not be confined near the interface, as charge could easily diffuse trough the bulk semiconductor. A structure of the kind $p-2D/i-n^{++}$, where $p$ and $n^{++}$ represent a p-type and a highly doped n-type semiconductor, respectively, could yield a higher efficiency than a standard $p-i-n$  structure, if an appropriate choice of the thickness of the i-Si is made. The thickness should be large enough to allow confinement and small enough not to increase significantly the overall resistance of the multilayer. 

\section{Conclusion}
A simple method to enhance the near-infrared absorption of intrinsic silicon was proposed and discussed in this paper. In particular, a 2D PtSe$_2$/i-Si bilayer was studied. It was shown that electron confinement in the 2D layer induces an hole-plasma in the silicon substrate. As a consequence, an upward bending of the valence band in the bulk semiconductor is induced, leading to an overall narrowing of the band-gap near the interface, as confirmed by XPS measurements. Furthermore, photo-induced Hall effect showed that an increase of absorbance by up to 20 times can be induced in Si, partially due to an extension of the absorbance spectrum to the infrared. The possibility of using such a strategy in solar cells technology was preliminarily discussed, too, and will be the subject of future works.

\section{Materials and Methods}\label{mat_met}
PtSe$_2$ samples films were purchased from \emph{SixCarbon Technology (Shenzhen)}. The average thickens of the layer, as measured by atomic force microscopy (AFM), was 1.8~nm. The 2D layers were grown on SiO$_2$ substrates and then transferred on the specific substrate (Si or $\alpha$-Al$_2$O$_3$) and annealed in vacuum at $150\,^{\circ}$C. The Raman spectrum in \figurename{~\ref{fig: Raman_2D_sapphire}} was measured with a \emph{WITec ALPHA300} (laser at 532~nm). A \emph{PerkinElmer} spectrometer was used to acquire the Fourier-transform infrared spectra shown in \figurename{~\ref{fig: FTIR}}. Pt thin films for photo-induced Hall effect measurements were deposited thorough \emph{Pulsed Laser Deposition (PLD)} method (laser energy: 300~mJ, frequency:  3~Hz , pulse width: 25~ns, deposition time: 180~s, pressure: $10^{-1}$~Torr). Magnetic fields were applied through an \emph{GMW~5403} electromagnet supplied by two \emph{RS-PRO~RSPD~3303C}. A halogen lamp \emph{Halopar~30} by \emph{Osram} was used as the light source and an \emph{FELH1250} edge-pass filter (long-pass wavelength at 1250~nm) by \emph{ThorLabs} was exploited to cut the visible part. The open-circuit Hall voltage was measured by using a \emph{Keithley~2182A} nanovoltmeter. Electric contacts to the samples were made by aluminum wire bondings.

\section*{Supporting Information}
Supporting Information is available after the references.

\section*{Acknowledgements}
This study was funded by the Science, Technology and Innovation Commission of Shenzhen Municipality (Project no. JCYJ20170307091130687). 

\bibliographystyle{iopart-num} 
\section*{References} 
\bibliography{bib_file}

\providecommand{\newblock}{}
\begin{thebibliography}{10}
\expandafter\ifx\csname url\endcsname\relax
  \def\url#1{{\tt #1}}\fi
\expandafter\ifx\csname urlprefix\endcsname\relax\def\urlprefix{URL }\fi
\providecommand{\eprint}[2][]{\url{#2}}

\bibitem{tiedje1984limiting}
Tiedje T, Yablonovitch E, Cody G~D and Brooks B~G 1984 {\em IEEE Trans.
  Electron Devices\/} {\bf 31} 711--716

\bibitem{richter2013reassessment}
Richter A, Hermle M and Glunz S~W 2013 {\em IEEE J. Photovolt.\/} {\bf 3}
  1184--1191

\bibitem{shalav2005application}
Shalav A, Richards B, Trupke T, Kr{\"a}mer K and G{\"u}del H~U 2005 {\em Appl.
  Phys. Lett.\/} {\bf 86} 013505

\bibitem{marti1996limiting}
Marti A and Ara{\'u}jo G~L 1996 {\em Sol. Energy Mater. Sol. Cells\/} {\bf 43}
  203--222

\bibitem{tobias2002ideal}
Tobias I and Luque A 2002 {\em Progress in Photovoltaics: Research and
  Applications\/} {\bf 10} 323--329

\bibitem{you2013polymer}
You J, Dou L, Yoshimura K, Kato T, Ohya K, Moriarty T, Emery K, Chen C~C, Gao
  J, Li G {\em et~al.\/} 2013 {\em Nat. Commun.\/} {\bf 4} 1446

\bibitem{meng2018organic}
Meng L, Zhang Y, Wan X, Li C, Zhang X, Wang Y, Ke X, Xiao Z, Ding L, Xia R {\em
  et~al.\/} 2018 {\em Science\/} {\bf 361} 1094--1098

\bibitem{lee2017review}
Lee T~D and Ebong A~U 2017 {\em Renewable Sustainable Energy Rev.\/} {\bf 70}
  1286--1297

\bibitem{devaney1990structure}
Devaney W, Chen W, Stewart J and Mickelsen R 1990 {\em IEEE Trans. Electron
  Devices\/} {\bf 37} 428--433

\bibitem{brabec2002low}
Brabec C~J, Winder C, Sariciftci N~S, Hummelen J~C, Dhanabalan A, van Hal P~A
  and Janssen R~A 2002 {\em Adv. Funct. Mater.\/} {\bf 12} 709--712

\bibitem{yan2010large}
Yan X, Cui X, Li B and Li L~s 2010 {\em Nano Lett.\/} {\bf 10} 1869--1873

\bibitem{tang2011infrared}
Tang J and Sargent E~H 2011 {\em Adv. Mater.\/} {\bf 23} 12--29

\bibitem{lin2009optical}
Lin C and Povinelli M~L 2009 {\em Opt. Express\/} {\bf 17} 19371--19381

\bibitem{kelzenberg2010enhanced}
Kelzenberg M~D, Boettcher S~W, Petykiewicz J~A, Turner-Evans D~B, Putnam M~C,
  Warren E~L, Spurgeon J~M, Briggs R~M, Lewis N~S and Atwater H~A 2010 {\em
  Nat. Mater.\/} {\bf 9} 239

\bibitem{novoselov2004electric}
Novoselov K~S, Geim A~K, Morozov S~V, Jiang D, Zhang Y, Dubonos S~V, Grigorieva
  I~V and Firsov A~A 2004 {\em Science\/} {\bf 306} 666--669

\bibitem{liu2008organic}
Liu Z, Liu Q, Huang Y, Ma Y, Yin S, Zhang X, Sun W and Chen Y 2008 {\em Adv.
  Mater.\/} {\bf 20} 3924--3930

\bibitem{chien2015graphene}
Chien C~T, Hiralal P, Wang D~Y, Huang I~S, Chen C~C, Chen C~W and Amaratunga
  G~A 2015 {\em Small\/} {\bf 11} 2929--2937

\bibitem{liu2015functionalized}
Liu Z, Lau S~P and Yan F 2015 {\em Chem. Soc. Rev.\/} {\bf 44} 5638--5679

\bibitem{wang2012electronics}
Wang Q~H, Kalantar-Zadeh K, Kis A, Coleman J~N and Strano M~S 2012 {\em Nat.
  Nanotechnol.\/} {\bf 7} 699

\bibitem{bernardi2013extraordinary}
Bernardi M, Palummo M and Grossman J~C 2013 {\em Nano Lett.\/} {\bf 13}
  3664--3670

\bibitem{jariwala2014emerging}
Jariwala D, Sangwan V~K, Lauhon L~J, Marks T~J and Hersam M~C 2014 {\em ACS
  Nano\/} {\bf 8} 1102--1120

\bibitem{mak2016photonics}
Mak K~F and Shan J 2016 {\em Nat. Photonics\/} {\bf 10} 216

\bibitem{tan2018emerging}
Tan C~L and Mohseni H 2018 {\em Nanophotonics\/} {\bf 7} 169--197

\bibitem{li2018photo}
Li D and Ruotolo A 2018 {\em Sci. Rep.\/} {\bf 8} 4372

\bibitem{ciarrocchi2018thickness}
Ciarrocchi A, Avsar A, Ovchinnikov D and Kis A 2018 {\em Nat. Commun.\/} {\bf
  9} 919

\bibitem{o2016raman}
O’Brien M, McEvoy N, Motta C, Zheng J~Y, Berner N~C, Kotakoski J, Elibol K,
  Pennycook T~J, Meyer J~C, Yim C {\em et~al.\/} 2016 {\em 2D Mater.\/} {\bf 3}
  021004

\bibitem{thomas1988infrared}
Thomas M~E, Joseph R~I and Tropf W~J 1988 {\em Appl. Opt.\/} {\bf 27} 239--245

\bibitem{druy1990fourier}
Druy M~A, Elandjian L, Stevenson W~A, Driver R~D, Leskowitz G~M and Curtiss L~E
  1990 Fourier transform infrared (ftir) fiber optic monitoring of composites
  during cure in an autoclave {\em Fiber Optic Smart Structures and Skins II\/}
  vol 1170 (International Society for Optics and Photonics) pp 150--160

\bibitem{tauc1966optical}
Tauc J, Grigorovici R and Vancu A 1966 {\em Phys. Status Solidi B\/} {\bf 15}
  627--637

\bibitem{viezbicke2015evaluation}
Viezbicke B~D, Patel S, Davis B~E and Birnie~III D~P 2015 {\em Phys. Status
  Solidi B\/} {\bf 252} 1700--1710

\bibitem{kandemir2018structural}
Kandemir A, Akbali B, Kahraman Z, Badalov S, Ozcan M, {\.I}yikanat F and Sahin
  H 2018 {\em Semicond. Sci. Technol.\/} {\bf 33} 085002

\bibitem{wang2015monolayer}
Wang Y, Li L, Yao W, Song S, Sun J, Pan J, Ren X, Li C, Okunishi E, Wang Y~Q
  {\em et~al.\/} 2015 {\em Nano Lett.\/} {\bf 15} 4013--4018

\bibitem{del1987modelling}
del Alamo J~A and Swanson R~M 1987 {\em Solid-State Electron.\/} {\bf 30}
  1127--1136

\bibitem{chang1970carrier}
Chang C and Sze S 1970 {\em Solid-State Electron.\/} {\bf 13} 727--740

\bibitem{cerofolini2003si}
Cerofolini G, Galati C and Renna L 2003 {\em Surf. Interface Anal.\/} {\bf 35}
  968--973

\bibitem{xpsHB}
Wagner C~D, Riggs W~M, Davis L~E, Moulder J~F and Muilenberg G~E 1979 {\em
  Handbook of X-ray photoelectron spectroscopy\/} (Perkin-Elmer Corporation)

\bibitem{persson2000plasma}
Persson C, Lindefelt U and Sernelius B 2000 {\em Solid-State Electron.\/} {\bf
  44} 471--476

\bibitem{balkanski1969infrared}
Balkanski M, Aziza A and Amzallag E 1969 {\em Phys. Status Solidi B\/} {\bf 31}
  323--330

\bibitem{schmid1981optical}
Schmid P 1981 {\em Phys. Rev. B\/} {\bf 23} 5531

\bibitem{alidoust2014observation}
Alidoust N, Bian G, Xu S~Y, Sankar R, Neupane M, Liu C, Belopolski I, Qu D~X,
  Denlinger J~D, Chou F~C {\em et~al.\/} 2014 {\em Nat. Commun.\/} {\bf 5} 4673

\bibitem{jain1991simple}
Jain S and Roulston D 1991 {\em Solid-State Electron.\/} {\bf 34} 453--465

\bibitem{molle2016electron}
Molle A, Lamperti A, Rotta D, Fanciulli M, Cinquanta E and Grazianetti C 2016
  {\em Adv. Mater. Interfaces\/} {\bf 3} 1500619

\end{thebibliography}

\newpage
\makeatletter
\renewcommand{\fnum@figure}{\figurename~S{\thefigure}}
\setcounter{figure}{0}
\makeatother
\section*{Supporting Information}
\textbf{Extending the Near-infrared Band-edge Absorption Spectrum of Silicon by Proximity to a 2D Semiconductor}
\\Valerio Apicella, Teslim Ayinde Fasasi, Antonio Ruotolo*
\newpage
\begin{figure}
	\includegraphics[width=1\columnwidth]{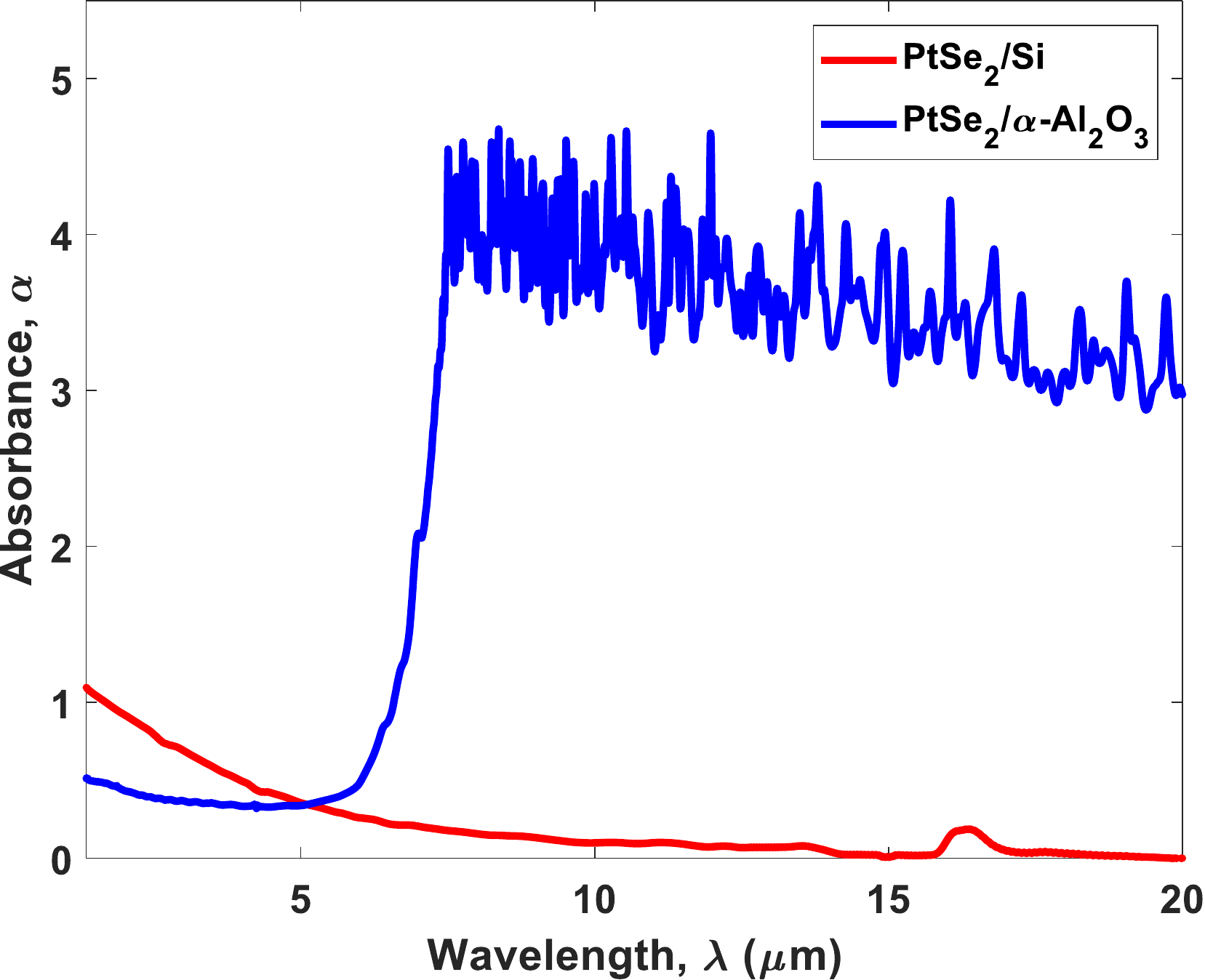}
	\caption{Full FTIR absorbance spectrum of PtSe$_2$/$\alpha$-Al$_2$O$_3$ and PtSe$_2$/Si.}
\end{figure}
The jump in the PtSe$_2$/$\alpha$-Al$_2$O$_3$ absorbance at 5000~nm is due to the sapphire substrate [1,2] and the signal becomes very noisy beyond such a wavelength. It is highlighted that the absorbance of infrared light in the PtSe$_2$/Si is higher than that in the 2D film for wavelengths as long as 5000~nm, despite the fact that Si is completely transparent to infrared for $\lambda>$1107~nm.
\newpage
\begin{figure}
	\includegraphics[width=1\columnwidth]{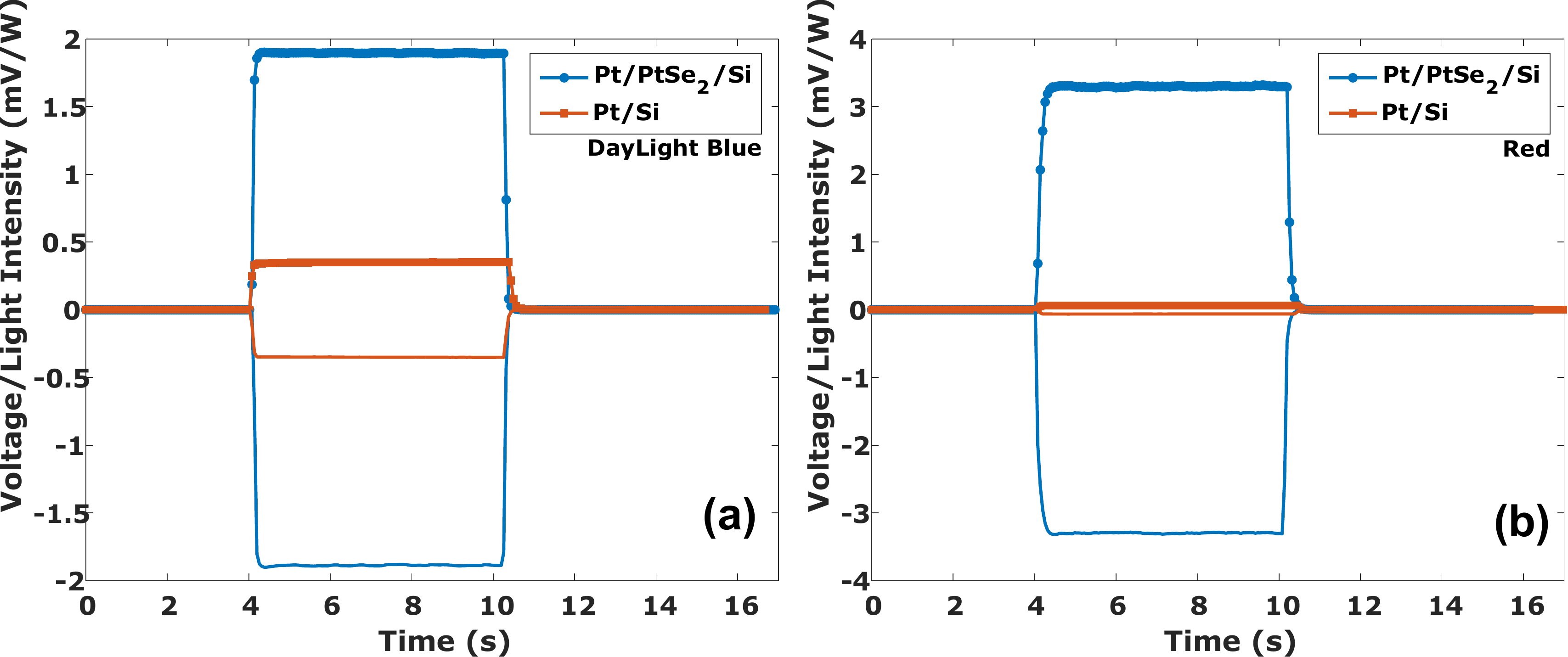}
	\caption{Comparison of photo-Hall effect for Pt/PtSe$_2$/Si and Pt/Si in daylight blue (a) and red light (b). Both positive and negative magnetic fields with 180~mT amplitude were applied and represented with marked and unmarked lines, respectively.}
\end{figure}
The light emitted by the halogen lamp was filtered in two different wavelength spectra. Two filters by Edmund Optics were used. The first one, namely a daylight blue filter, has cut-on and cut-off wavelengths at 340 and 465~nm, respectively. The second one, a red filter, has the cut-on wavelength at 600~nm. It is possible to observe that the peak of photo-induced Hall voltage shifts toward longer wavelengths when the 2D semiconductor exerts its proximity on the Si substrate.
\newpage
\begin{figure}
	\includegraphics[width=1\columnwidth]{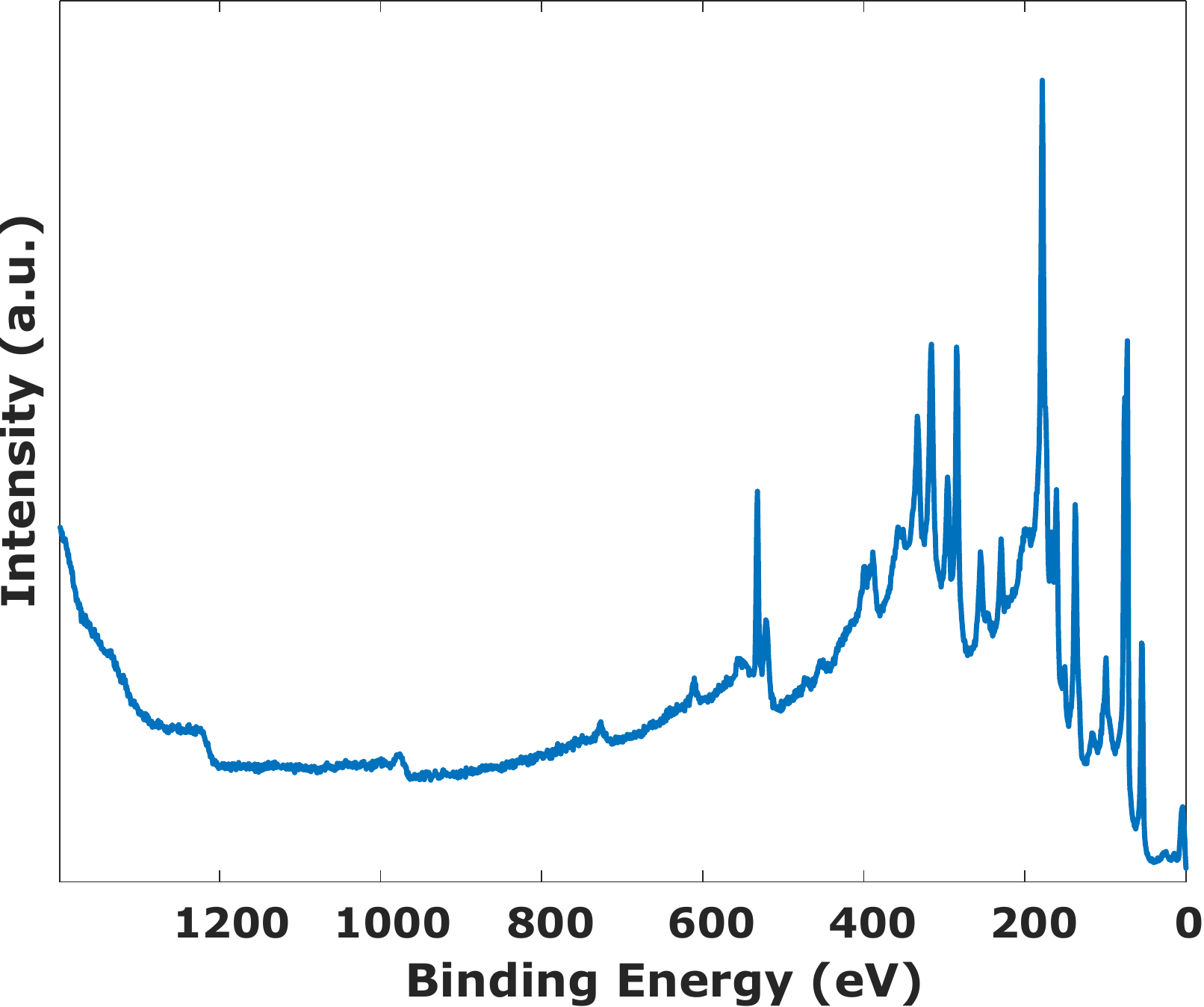}
	\caption{XPS spectrum as measured on the PtSe$_2$/Si sample.}
\end{figure}
-
\newpage
\begin{figure}
	\includegraphics[width=1\columnwidth]{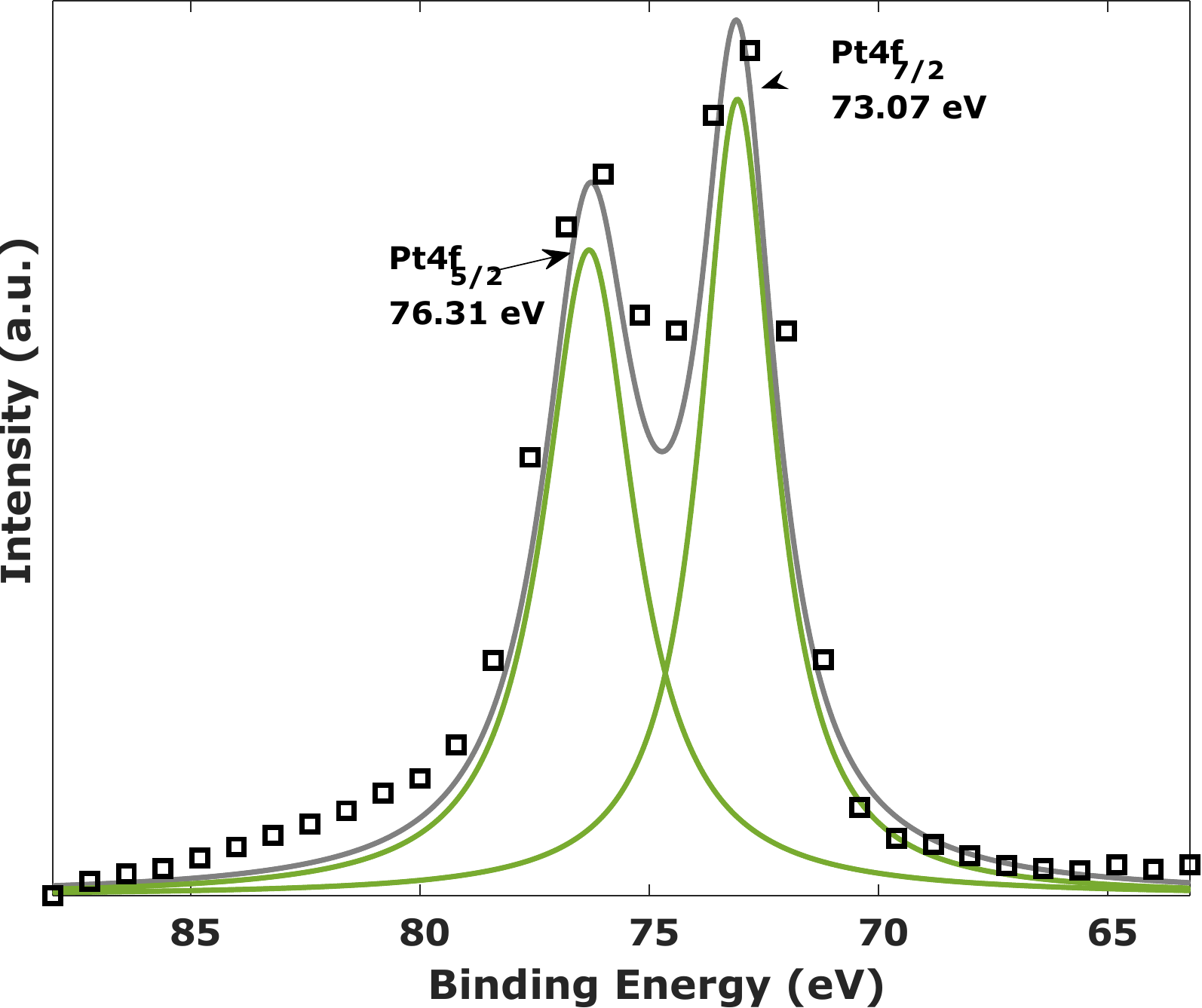}
	\caption{Deconvolution of the Pt XPS peaks. The two peaks at 73.07~eV and 76.31~eV correspond to Pt4f$_{7/2}$ and Pt4f$_{5/2}$, respectively [3,4].}
\end{figure}
-
\newpage
\begin{figure}
	\includegraphics[width=1\columnwidth]{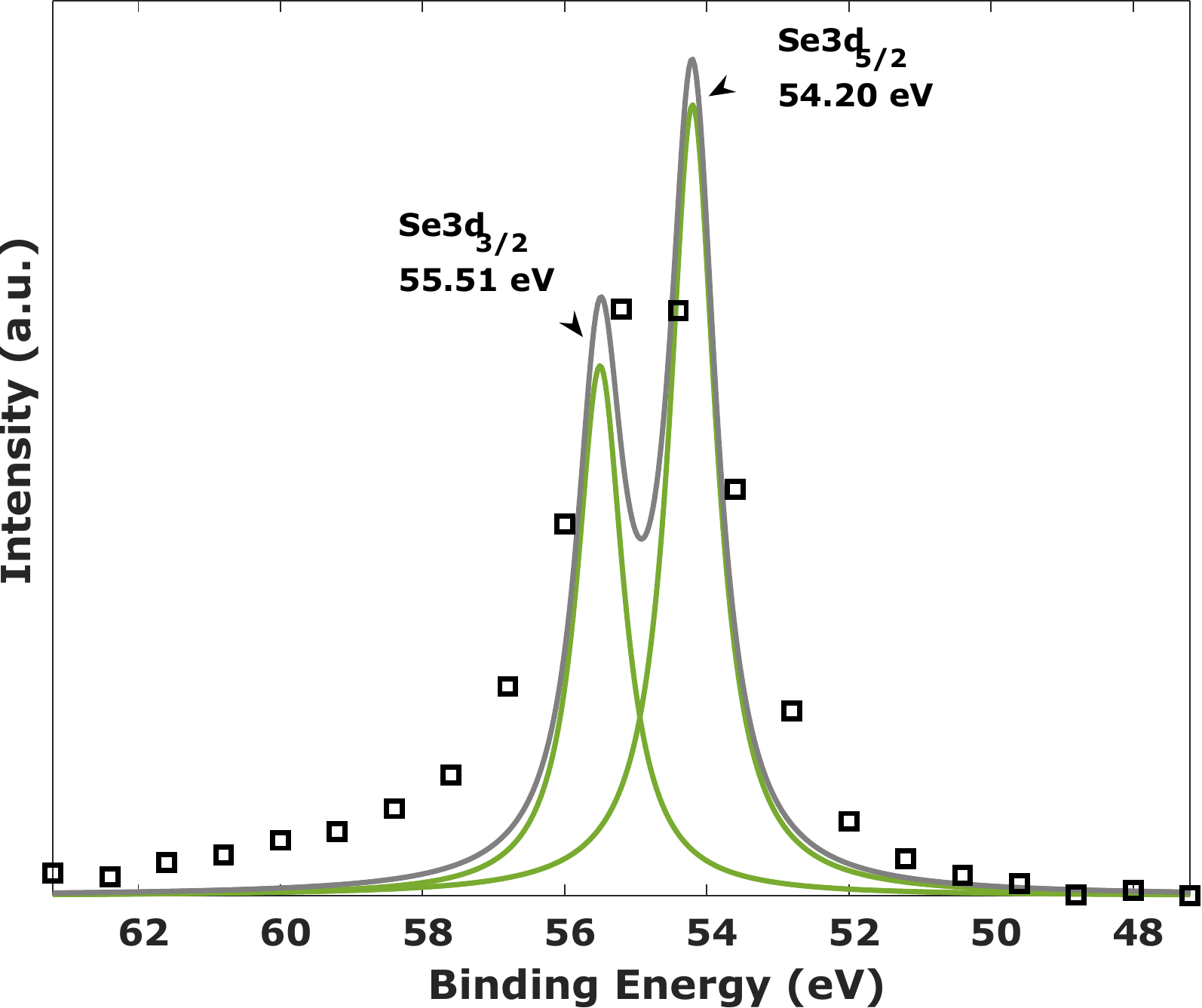}
	\caption{Deconvolution of the Se XPS peaks. The two peaks at 54.20~eV and 55.51~eV correspond to Se3d$_{5/2}$ and Se3d$_{3/2}$, respectively [3,4].}
\end{figure}
The position of peaks for Pt and Se is in agreement with XPS measurements on PtSe$_2$ reported in refs. [3,4].
\newpage
\section*{References} 
$[1]$ M. E. Thomas, R. I. Joseph, W. J. Tropf, Applied Optics 1988, \textbf{27}, 2 239.\\
$[2]$ M. A. Druy, L. Elandjian, W. A. Stevenson, R. D. Driver, G. M. Leskowitz, L. E. Curtiss, In Fiber Optic Smart Structures and Skins II, volume 1170. International Society for Optics and Photonics 1990 150-160.\\
$[3]$ C. Xie, L. Zeng, Z. Zhang, Y.-H. Tsang, L. Luo, J.-H. Lee, Nanoscale 2018, \textbf{10}, 32 15285.\\
$[4]$ L.-H. Zeng, S.-H. Lin, Z.-J. Li, Z.-X. Zhang, T.-F. Zhang, C. Xie, C.-H. Mak, Y. Chai, S. P. Lau, L.-B. Luo, et al., Advanced Functional Materials 2018, \textbf{28}, 16 1705970.

\end{document}